\documentclass[12pt]{article}
\usepackage{epsfig,axodraw,graphicx,psfrag}

\renewcommand{\thefootnote}{\fnsymbol{footnote}}

\newcommand{\prepr}[1] {\begin{flushright}  {\bf #1} \end{flushright} \vskip 1.cm}
\newcommand{\titul}[1] {\boldmath \begin{center}{\Large {\bf #1 } } \end{center}
\vskip 0.8cm}
\newcommand{\autor}[1] {\begin{center}  {\bf \lineskip .3cm #1  }
                        \end{center} }
\newcommand{\lugar}[1] {\begin{center}  {\normalsize \bf \it #1   } \end{center}}

\newcounter{muni}

\begin{document}
\hbadness=10000
\pagenumbering{arabic}
\begin{titlepage}

\prepr{hep-ph/0103008\\
\hspace{30mm} KEK-TH-744 \\
\hspace{30mm} IPAS-HEP-2001-02 \\
\hspace{30mm} February 2001}

\begin{center}
\titul{\bf Effect of Supersymmetric phases
on the Direct CP Asymmetry of $B\to X_d\gamma$}

\autor{A.G. Akeroyd$^{\mbox{1}}$\footnote{akeroyd@post.kek.jp },
Y.-Y. Keum$^{\mbox{2}}$\footnote{keum@phys.sinica.edu.tw } and
S. Recksiegel$^{\mbox{1}}$\footnote{stefan@post.kek.jp} }

\lugar{ $^{1}$ Theory Group, KEK, Tsukuba, Ibaraki 305-0801, Japan }
\lugar{ $^{2}$ Institute of Physics, Academia Sinica,\\
Nankang, Taipei, Taiwan , 11529, R.O.C.}

\end{center}

\vskip2.0cm

\begin{abstract}
\noindent{ We investigate the effect of supersymmetric CP violating phases on 
the inclusive decay $B\to X_d\gamma$. Although such a decay
contains a large background from $B\to X_s\gamma$, if isolated 
it may exhibit sizeable CP violation, 
both in the Standard Model (SM) and in the context of models beyond
the SM. With unconstrained supersymmetric CP violating phases
we show that the direct CP asymmetry 
(${\cal A}_{CP}$) lies in the region $-40\% \le {\cal A}_{CP}\le 
40\%$, where a positive asymmetry would constitute a clear 
signal of physics beyond the SM. Even if a direct 
measurement of $B\to X_d\gamma$ proves too difficult experimentally,
its asymmetry contributes non-negligibly to the measurements 
of ${\cal A}_{CP}$ for $B\to X_s\gamma$, and thus should be included
in future analyses. We show that there may be both constructive and
destructive interference between  ${\cal A}^{d\gamma}_{CP}$ and
${\cal A}^{s\gamma}_{CP}$.}
\end{abstract}

\end{titlepage}
\thispagestyle{empty}
\newpage

\pagestyle{plain}
\renewcommand{\thefootnote}{\arabic{footnote} }
\setcounter{footnote}{0}

\section{Introduction}

Theoretical studies of rare decays of $b$ quarks have attracted
increasing attention with the recent turn-on of the $B$ factories at
KEK and SLAC \cite{Bphys},\cite{Plen}. In this paper we are concerned 
with the rare decay $b\to d\gamma$ which proceeds via an electromagnetic
penguin diagram, and is sensitive to the CKM matrix element 
$V_{td}$. The latter is so far unmeasured directly and in the 
Wolfenstein parameterisation is given by: 
\begin{equation}
V_{td}=A\lambda^3(1-\rho-i\eta)
\end{equation}
By assuming unitarity of the CKM matrix and using various experimental data 
($\epsilon_K,|V_{ub}/V_{cb}|,$
$\Delta m_d,\Delta m_s$)
an allowed region for $\rho$ and $\eta$ at 95\% c.l.\ \cite{CKM} can be
obtained. This enables one to
infer the range of $V_{td}$ consistent with the unitarity of the
CKM matrix.

A direct measurement of $V_{td}$ is thus desirable and we will be 
considering the inclusive decay $B\to X_d\gamma$.
Current measurements of $B^0_d-\overline {B^0_d}$ mixing 
yield $0.0065\le |V_{td}V^*_{tb}|\le 0.010$ \cite{EPJ8}, where
the main source of error is the theoretical uncertainty in the
hadronic matrix element $f_{B_d}\sqrt{B_{B_d}}$. 
Alternative ways to measure the ratio $|V_{td}/V_{ts}|$
have been proposed and include the 
$B^0\overline {B^0}$ mixing ratio $\Delta m_d/\Delta m_s$
and the ratio of the branching ratios (BR)
of $B\to X_sl^+l^-$ and $B\to X_dl^+l^-$
\cite{EPJ8}.

The short distance contribution to $B\to X_d\gamma$
is the quark transition $b\to d\gamma$. Estimates
for the long distance contributions can be found in 
\cite{Ldist, Donoghue:1997fv}
and their relative size is expected to be around $10\%$ 
of the short distance contribution. 
Experimental upper limits exist
for the branching ratios (BRs) of the exclusive decay channels, 
$B\to \rho^0\gamma$ and $B\to \rho^+\gamma$. 
CLEO \cite{pyCLEO} obtains $\le 1.7\times 10^{-5}$ and
$\le 1.3\times 10^{-5}$ respectively, with corresponding measurements
by BELLE \cite{pyKEK} of
$\le 0.56\times 10^{-5}$ and $\le 2.27\times 10^{-5}$.

There is considerable motivation for calculating the 
BR and CP asymmetry (${\cal A}_{CP}$) of the inclusive channel
$B\to X_d\gamma$:

\begin{itemize}

\item[{(i)}] It provides a theoretically clean way of measuring
$V_{td}$, as proposed in \cite{Ali92, plb429}

\item[{(ii)}] ${\cal A}_{CP}$ in the SM is sizeable, and 
much larger than that for $b\to s\gamma$ \cite{plb429}.

\item[{(iii)}]$ {\cal A}_{CP}$ is sensitive
to new physics which contributes to $C_7$ 
\cite{plb399,plb460,AAY,KSW}.

\item[{(iv)}] The current measurement of ${\cal A}_{CP}$
for $b\to s\gamma$ by the CLEO Collaboration \cite{CLEO} is
sensitive to events from $b\to d\gamma$. Therefore knowledge of
${\cal A}_{CP}$ for $b\to d\gamma$ is essential, in order to
compare experimental data with the theoretical prediction 
in a given model.

\end{itemize}

We are interested in the effect of unconstrained supersymmetric (SUSY)
CP violating phases on the inclusive decay BR$(B\to X_d\gamma)$. 
We will be working in the context of the effective SUSY model proposed 
in \cite{effSUSY}. Such a model allows one to consider the full 
impact of the phases on the rare decays of $B$ mesons, while simultaneously
satisfying the stringent bounds on the Electric Dipole Moments 
of the electron and neutron. Such phases
may be crucial for generating the observed matter-antimatter 
asymmetry in the universe \cite{barygen}.

Our work is organized as follows. In section 2 we introduce the
decays $b\to d\gamma$ and $b\to s\gamma$. In section 3 we outline 
our approach to calculate the CP asymmetries, while section 4 presents
the numerical results. Finally, section 5 contains our conclusions.

\boldmath
\section{The decays $b\to d\gamma$ and $b\to s\gamma$}
\unboldmath

Much theoretical study has been devoted to the decay $b\to s\gamma$ 
due to its sensitivity to physics beyond the SM
\cite{bsytheory}. Exclusive channels ($B\to K^*\gamma$ etc.) 
and the inclusive channel have been measured at CLEO, ALEPH, BELLE and 
BaBar \cite{bsymeasure}.
The related decay $b\to d\gamma$ has received less attention
although is expected to be observed at the $B$ factories, at least
in some exclusive channels. 

Ref. \cite{plb429} calculated BR$(B\to X_d\gamma$)
in the context of the SM. It was shown that the ratio $R$ defined by
\begin{equation}
R={BR(B\to X_d\gamma)\over {BR(B\to X_s\gamma)}}
\end{equation}
is expected to be in the range $0.017 < R < 0.074$, corresponding
to BR$(B\to X_d\gamma)$ of order $10^{-5}\to 10^{-6}$. 
With $10^8$ $B\overline B$ pairs expected
from the $B$ factories, one would be able to produce $10^2\to 10^3$
$b\to d\gamma$ transitions.  In the ratio $R$ most of the theoretical
uncertainties cancel, and hence $R$ may provide a theoretically clean
way of extracting the ratio $|V_{td}/V_{ts}|$.

The CP asymmetry (${\cal A}_{CP}$), defined by\footnote{
Note that our definition ${\cal A}_{CP}$ contains a relative
minus sign compared to that used in \cite{plb429},\cite{plb460},
\cite{CLEO}}
\begin{equation}
{\cal A}^{d\gamma(s\gamma)}_{CP}={{\Gamma(\overline B\to X_{d(s)}\gamma)-
\Gamma(B\to X_{\overline d(\overline s)}\gamma)}
\over {\Gamma(\overline B\to X_{d(s)}\gamma)+\Gamma(B\to X_{\overline
d(\overline s)}\gamma)}}={\Delta\Gamma_{d(s)}\over \Gamma^{tot}_{d(s)}}
\label{ACPdef} \end{equation}
is expected to lie in the range $-7\%\le {\cal A}_{CP}^{d\gamma}\le -35\%$ 
in the SM \cite{plb429}, where the uncertainty arises from 
varying $\rho$ and $\eta$
in their allowed regions. Also included is the scale dependence $(\mu_b)$ 
of ${\cal A}_{CP}^{d\gamma}$ which occurs from varying
$m_b/2\le \mu_b \le 2m_b$. For definiteness we fix
$\mu_b=4.8$ GeV, and find $-5\%\le A_{CP}^{d\gamma}\le -28\%$.
Therefore  ${\cal A}_{CP}^{d\gamma}$ is much larger than
${\cal A}_{CP}^{s\gamma}$ ($\le 0.6\%$). By estimating values for detection 
efficiencies, it has been argued in \cite{KSW}
that ${\cal A}^{d\gamma}_{CP}$ may be statistically more accessible than
${\cal A}^{s\gamma}_{CP}$, at least in the context of the SM.
This analysis assumes that 
$B\to X_d\gamma$ can be clearly isolated from $B\to X_s\gamma$.

However, it is known that isolating the signal $B\to X_d\gamma$ would be
an experimental challenge since $B\to X_s\gamma$ constitutes a
serious background. \cite{KSW} has suggested
several ways to overcome this problem, e.g. demanding a 
higher energy cut on $\gamma$, since $\gamma$ from 
$B\to X_d\gamma$ will be more energetic than that from 
$B\to X_s\gamma$.
Energy cuts can be used to separate $b\to s\gamma$
events from charmed background since there is a
high photon energy region that is inaccessible to
charmed states because of the mass of the charm quarks.
This method is not feasible for extracting $b\to d\gamma$
events from a $b\to s\gamma$ sample. Although the strange
quark mass is larger than the down quark mass, the respective lightest
hadronic single particle final states, $K^*$ and $\rho$, have almost 
the same mass (they actually overlap strongly). The lightest multi-particle
states are $K\pi$ and $\pi\pi$, respectively, but even here
effects such as bound state effects (neglected in \cite{KSW})
smear the spectra out over regions of the order of 200 MeV.
These effects constitute one of the major theoretical
uncertainties in the extraction of $BR(B\to X_s\gamma)$ from
the measured part of the spectrum, and they make a separation
of $b\to d\gamma$ and $b\to s\gamma$ via energy cuts impossible.
A comparison of the photon energy spectra for $b\to s\gamma$
and $b\to d\gamma$ was made in \cite{Ali92}, and showed that
the photon spectra for both decays are very similar.

A more promising approach constitute exclusive channels 
\cite{Bphys,Alitalk,Ali}.
The improved $K/\pi$ separation at the $B$ factories may enable
the inclusive $B\to X_d\gamma$ decay to be reconstructed by summing 
over the relevant exclusive channels 
as done by CLEO in the measurement of $B\to X_s\gamma$. \cite{KSW} 
suggested using a semi-inclusive sample of 
$B \to \gamma + n\pi$ decays with a maximum of $n$ (say 5) mesons 
together with a corresponding measurement of
$B \to \gamma+K+ (n-1)\pi$. The ratio of the widths 
of the semi-inclusive samples would enable the
total inclusive rate to be deduced to a very good approximation.
Although the extraction of the branching ratio for $b\to d\gamma$
from exclusive channels might suffer additional uncertainties
with respect to $b\to s\gamma$ \cite{Donoghue:1997fv}
the asymmetry should not be affected by these.

If ${\cal A}^{d\gamma}_{CP}$ and ${\cal A}^{s\gamma}_{CP}$ 
cannot be separated, then only their sum can be measured.
In the context of the SM (with $m_s=m_d=0$)
the unitarity of the CKM matrix ensures that the sum
is zero \cite{neubert,soares}.
This relation holds only for the
short distance contribution, which is expected to
be dominant (c.f.\ introduction).
In the presence of new physics such a cancellation does not
occur, as will be
shown in section 4. As stressed earlier, a 
reliable prediction of 
${\cal A}^{d\gamma}_{CP}$ in a given model is necessary 
since it contributes to the measurement of ${\cal A}^{s\gamma}_{CP}$.
The CLEO result is sensitive to a weighted sum of CP asymmetries, 
given by:
\begin{equation}
{\cal A}^{exp}_{CP}=0.965{\cal A}^{s\gamma}_{CP}+0.02{\cal A}^{d\gamma}_{CP}
\label{CLEOeq} \end{equation}  
The latest measurement stands at $-27\% < {\cal A}^{exp}_{CP} < 10\%$ 
(90\% c.l.) \cite{CLEO}. 
The small coefficient of ${\cal A}^{d\gamma}_{CP}$ is caused by
the smaller BR$(B\to X_d\gamma)$ (assumed to be $1/20$ that
of BR$(B\to X_s\gamma)$) and inferior detection efficiencies, but 
may be partly compensated by the larger value for ${\cal A}^{d\gamma}_{CP}$.
We shall see that there can be both constructive and destructive
interference between the two terms in eq.~~(\ref{CLEOeq}). These effects will
be especially important for measurements in future high luminosity  
runs of $B$ factories, in which the precision is expected to reach
a magnitude where the $b\to d\gamma$ contribution becomes crucial.
For integrated luminosities of 200 fb$^{-1}$ (2500 fb$^{-1}$) \cite{alex}
anticipates a precision of $3\%(1\%)$ in the measurement of
${\cal A}^{exp}_{CP}$.

\section{Direct CP Asymmetry in $B \to X_{d,s} \gamma$}

In this paper we explore the effect of CP violating SUSY phases
on the direct CP asymmetry of the inclusive decay $B \to X_{d(s)} \gamma$.
We will show that the asymmetry ${\cal A}_{CP}^{d\gamma}$
may be quite different from the SM prediction
in a wide region of parameter space consistent with experimental
bounds from the Electric Dipole Moment (EDM) and BR($B \to X_{s}\gamma$).

In our analysis we adapt the ``effective SUSY'' model, proposed in
\cite{effSUSY}. This model permits unrestricted SUSY phases
and evades the electric dipole moment (EDM) constraint by invoking
large masses (of order 20 TeV) for the first two generations of
sfermions, thus maintaining their contribution to the EDMs within
the experimental limits. The third generation sfermions 
are allowed to be relatively light ($\le 1$ TeV) with large
phases in $\mu$ and the soft breaking term $A_t$.
Such an approach has been used on several occasions \cite{baek},
\cite{Romao} in order to study the maximum impact of SUSY phases 
on rare $B$ decays.
It was pointed out in \cite{PRL82} that the third generation squarks
may contribute to the EDMs via non--negligible two loop diagrams,
and we include this constraint in our analysis. 

Previous work on the magnitude of ${\cal A}^{d\gamma}_{CP}$ in models
beyond the SM include multi--Higgs doublet models 
\cite{AAY, KSW}, a Left-Right symmetric model 
\cite{plb399} and the MSSM without SUSY phases \cite{plb460}. 
A common feature to
all these analyses is the possibility of ${\cal A}^{d\gamma}_{CP}$
being of opposite sign to that of the SM, which would be
a clear signal of new physics.
Ref.~\cite{plb460} took $A_t$ and $\mu$ to be real, 
and thus the only source of CP violation was the CKM phase. 
They found two phenomenological acceptable regions, corresponding
to $2\% \le {\cal A}^{d\gamma}_{CP}\le 21\%$ and 
$-45\% \le {\cal A}^{d\gamma}_{CP}\le -5\%$.
It is instructive to consider the impact of unconstrained
SUSY phases on the inclusive decay $B\to X_d\gamma$ by
taking $A_t$ and $\mu$ complex. The same approach has been used in 
Ref.\cite{baek}, and it was shown that ${\cal A}^{s\gamma}_{CP}$ 
may lie in the range $-16\% \le {\cal A}^{s\gamma}_{CP}\le 16\%$.   
Other authors \cite{bsy} have considered a variety of SUSY models with
additional theoretical assumptions, resulting in lower values
for the maximum value of ${\cal A}^{s\gamma}_{CP}$. 
The study of $b\to d\gamma$ in these models
will be considered in future work \cite{progress}.

We now briefly outline our approach for the calculation
of $A^{d(s)\gamma}_{CP}$. 
We assume that flavour changing neutral current
vertices induced by the gluino and neutralino are absent.
Therefore to lowest order, the decay at quark level proceeds via the
following diagrams, where the photon may be emitted from any
charged line: 

\vspace{10mm}

\begin{center}
\vspace{-50pt} \hfill \\
\begin{picture}(200,70)(0,25) 
\ArrowLine(-20,25)(40,25)
\ArrowLine(40,25)(80,25)
\ArrowLine(80,25)(120,25)
\ArrowLine(120,25)(180,25)
\Photon(80,25)(80,65){4}{8}
\DashCArc(80,25)(40,180,0){5}
\Vertex(80,25){2}
\Text(10,35)[]{$b$}
\Text(150,-4)[]{$W^{\pm},H^{\pm},\chi^{\pm}$}
\Text(95,57)[]{$\gamma$}
\Text(150,35)[]{$s,d$}
\Text(60,35)[]{$t$,$\tilde{t}$}
\end{picture}
\end{center}

\vspace{20mm}

The effective  Hamiltonian for $b \to d \gamma$ is given by
\begin{equation}
{\cal H}_{eff} = -{4 G_F \over \sqrt{2}} V_{td}^{*} V_{tb} \sum_{i=1}^{8}
C_{i}(\mu_b) Q_{i}(\mu_b)
\label{Heff}
\end{equation}
where $Q_{i}(\mu_b)$ is the current density operator for the $\triangle B =1$
transition and $C_{i}(\mu_b)$ is its Wilson coefficient.
The relevant operators for $b \to d \gamma$ decay are given by
\begin{eqnarray}
Q_{2} &=& \bar{d}_{L} \gamma^{\mu}c_{L} \bar{c}_{L} \gamma^{\mu} b_{L}, 
\nonumber \\
Q_{7} &=& {e \over 16 \pi^2} m_b \bar{d}_{L} \sigma^{\mu\nu}b_{R}F_{\mu\nu},
\label{op} \\
Q_{8} &=& {g_s \over 16 \pi^2} m_b \bar{d}_{L} \sigma^{\mu\nu} T^{a}
b_{R}G^{a}_{\mu\nu}. \nonumber 
\end{eqnarray}
The analogous formulae for the $b 
\to s\gamma$ decay can obtained 
from Eqs.(\ref{Heff}) and (\ref{op}) by making the
replacement $d\to s$.

The asymmetry ${\cal A}_{CP}^{d(s)\gamma}$ can be written as:
\begin{eqnarray}
{\cal A}_{CP}^{d(s)\gamma}&=&{10^{-2}\over |C_7|^2}
\Big(1.17\times {\rm Im}[C_2C_7^*]-9.51\times
{\rm Im}[C_8C_7^*]+0.12\times {\rm Im}[C_2C_8^*]\ \nonumber \\
&& \hspace{20mm} -9.40\times {\rm Im}[\epsilon_{d(s)}
C_2(C_7^*-0.013C_8^*)]\Big) \nonumber \\
\cr
&=& {1.1 \over (1 + Re[\xi_7] )^2 + (Im[\xi_7])^2} 
\left[ 0.54 \,\, Im[\xi_7] - 0.25 \,\, Im[\xi_8] 
- 0.19 \,\, Im[\xi^{*}_7 \xi_8] \right. 
\nonumber \\
\cr
&& \hspace{20mm} 
\left. + 3.21 \,\, Im[\epsilon_{s(d)} (1 + 0.65 \xi_7^{*} + 0.04 \xi_8^{*})]
\right]
\label{Acp}
\end{eqnarray}
where $\xi_{7,8} = (C_{7,8} - C_{7,8}^{SM})/C_{7,8}^{SM}$.
$C_{7,8}$ include the total contribution while $C_{7,8}^{SM}$ contain
only the Standard Model one. 
Here we use $C_7^{SM} = -0.30$ and $C_8^{SM} = -0.14$.

In the SM since all the Wilson coefficients are real,
the only contribution comes from the final term, 
which corresponds to the CKM phase in 
$\epsilon_{x} = V_{ux}^{*}V_{ub}/V_{tx}^{*}V_{tb}$.

The branching ratios in terms of the new physics contributions are given by
\begin{eqnarray}
{\bf Br}(B \to X_s \gamma) &=& {|V^{*}_{ts} V_{tb}|^2 \over |V_{cb}|^2} \,\,
{6 \alpha_{em} \over \pi f(z)} \,\, |C_7|^2 {\bf Br}(B\to X_c e \bar{\nu}_e),
\nonumber \\
\cr
&\simeq& (3.48 \pm 0.31) \times 10^{-4} \,\, 
\left[(1 + Re[\xi_7])^2 + (Im[\xi_7])^2 \right], \label{Brsg} \\
\cr
{\bf Br}(B \to X_d \gamma) &=& \lambda^2 \,\, [(1-\rho)^2 + \eta^2] \,\,
{\bf Br}(B \to X_s \gamma).
\label{Brdg}
\end{eqnarray}

The dominant contribution to the decay comes from $C_7$ evaluated
at the scale $m_b$, which may be divided into contributions from  
$W^\pm$,$H^\pm$ and $\chi^\pm$ respectively:
\begin{equation}
C_7=C_7^{W}+C_7^{H}+C_7^{\chi}
\end{equation}
Both $C_7^{W}$ and $C_7^{H}$ are purely real, while 
$C_7^{\chi}$ may possess an imaginary part. We will use
the leading order expressions for $C_7$ which may be found
in \cite{Buras:1994xp}.
Although higher order corrections to $C_7^W$ are available, 
the corrections to $C_7^{H}$ and $C_7^{\chi}$ are only valid in
certain limiting cases which are not generally applicable 
\cite{bsylim}. Therefore to
be consistent to a given order we limit ourselves to the leading
order expressions for the Wilson coefficients. Such an approach has 
also been adopted in \cite{KSW} and \cite{baek}. 
The magnitude of $|C_7|$ is constrained by measurements of the
branching ratio of $b\to s\gamma$, and so we only consider points
in parameter space that satisfy $0.2 \le |C_7(m_b)| \le 0.38$.

In effective supersymmetry the main contribution to the
EDMs of the electron and neutron comes from the two-loop constraint induced by
Barr-Zee diagrams involving the Higgs pseudoscalar ($A^0$)
and the third generation squarks \cite{PRL82}. 
Since the neutron EDM has large hadronic uncertainties,
we only consider the upper bound of the electron EDM in our analysis.
For large values of $\tan\beta$ this contribution may exceed the
present experimental bounds. We have checked that the region
$\tan\beta\le 30$ comfortably satisfies this EDM constraint
even for a light $\tilde t_1$ ($\le 200$ GeV). In particular,
larger values of the Higgs pseudoscalar mass ($M_A\ge 400$ GeV)
induce enough suppression to allow a sizeable parameter space
with $\tan\beta\ge 30$. We shall see that the full
range of values for ${\cal A}_{CP}^{d\gamma}$ and ${\cal A}_{CP}^{s\gamma}$
can be obtained for any value of $\tan\beta\ge 10$, and so such a 
constraint has little impact on our results.


\section{Numerical Results}
We vary the (SUSY) parameters in the following range:\\

\begin{tabular}{|c|c|c|c|c|c|c|c|c|c|c||c|c|}
\hline
&$M$&$M'$&$\tan\beta$&$m_{H^\pm}$&$M_Q$&$M_U$&$\mu$&$\phi_\mu$&$A_t$&$\phi_A$&$\rho$&$\eta$\\
\hline
min&0&0&1&200&0&0&0&0&0&0&-0.1&0.2\\
max&400&400&30&500&200&200&200&$2\pi$&300&$2\pi$&0.4&0.5\\
\hline
\end{tabular}\\

\vspace*{0.3cm}
Here $M$ and $M'$ are respectively the $SU(2)$ and $U(1)$ gaugino 
soft masses; $A_t$ is the tri-linear soft mass for  $\tilde t$,
with phase $\phi_A$; $M_Q$ and $M_U$ are the soft masses
for the third generation squarks.
We respect the direct search lower limits on the masses of $\tilde t_1$,
$\chi^{\pm}$ by discarding generated points that do not
pass our cuts of $m_{\tilde t_1} > 90$ GeV and $m_{\chi^{\pm}_1} > 80$ GeV
in addition to the cut on $C_7$ mentioned in Sec.~3.
We vary $\rho$ and $\eta$ in the range allowed by present 
CKM fits for the SM \cite{CKM}. Note that in the effective SUSY
model one should strictly only include the constraint from
$|V_{ub}/V_{cb}|$, which corresponds to varying 
$\rho$ and $\eta$ in a semi-circular band in the 
$\rho-\eta$ plane. This enlarged parameter space has little effect
on our graphs, except for Fig.3, which will be commented on below.

\begin{figure}
\begin{center}
\psfrag{XXX}{$m_{\tilde t_1}$}  \psfrag{YYY}
 {${\cal A}^{d\gamma}_{CP}$}
\includegraphics[width=10cm]{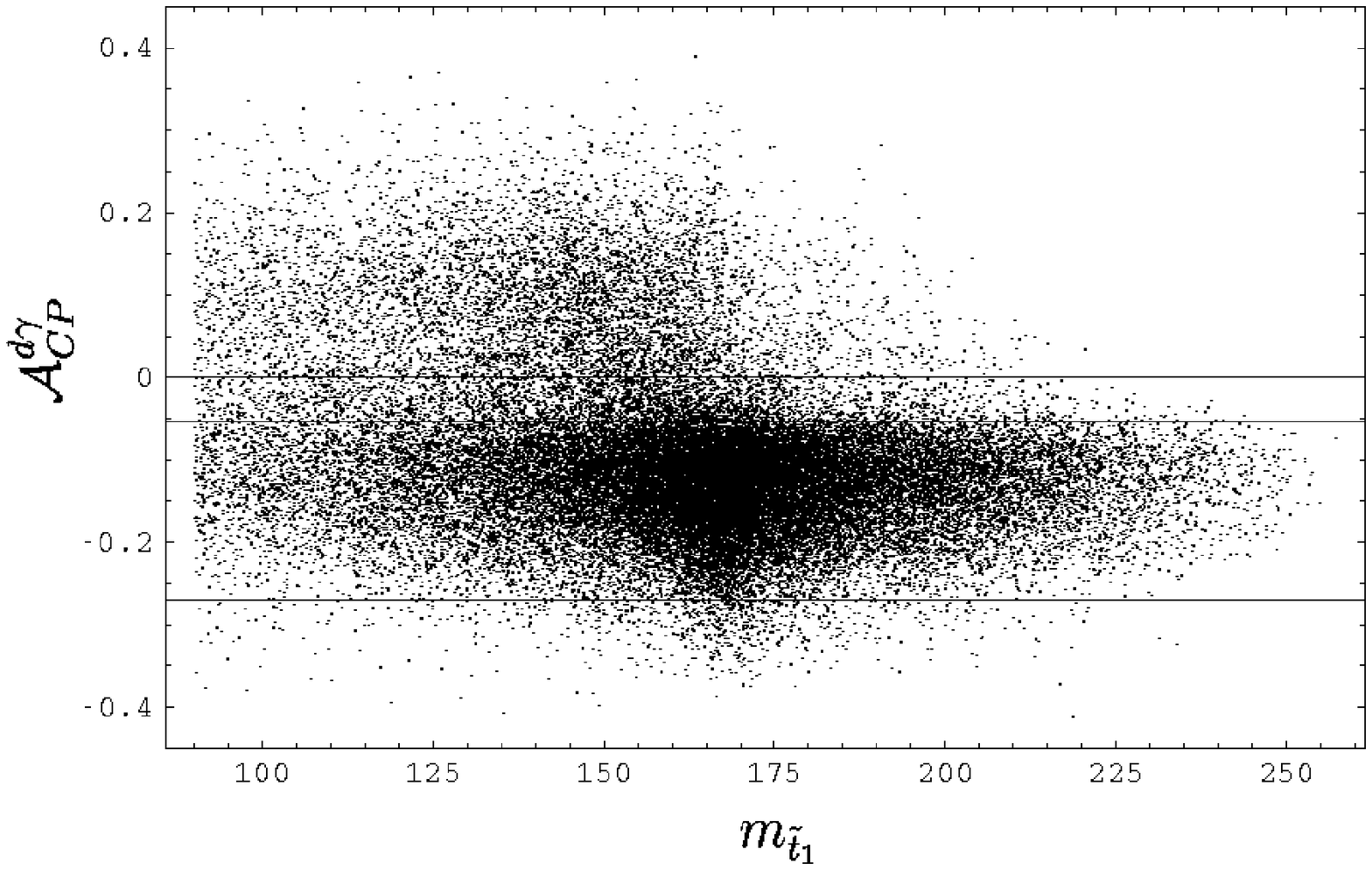}
\end{center}
\vspace*{-.5cm}
\caption{${\cal A}^{d\gamma}_{CP}$ against $m_{\tilde t_1}$}
\label{mst1}
\end{figure}

If the signal for the inclusive decay can be isolated
then a positive asymmetry would be a clear sign of new physics. 
In Fig.~\ref{mst1} we plot ${\cal A}^{d\gamma}_{CP}$ against 
$m_{\tilde t_1}$, which clearly shows that a light $\tilde t_1$ may 
drive ${\cal A}_{CP}^{d\gamma}$ positive, reaching maximal 
values close to $+40\%$. For $\tilde t_1$ heavier than 250 GeV
the ${\cal A}_{CP}^{d\gamma}$ lies within the SM range, which
is indicated by the two horizontal lines.

\begin{figure}
\begin{center}
\psfrag{XXX}{$\tan\beta$}  \psfrag{YYY}
 {${\cal A}^{d\gamma}_{CP}$}
\includegraphics[width=10cm]{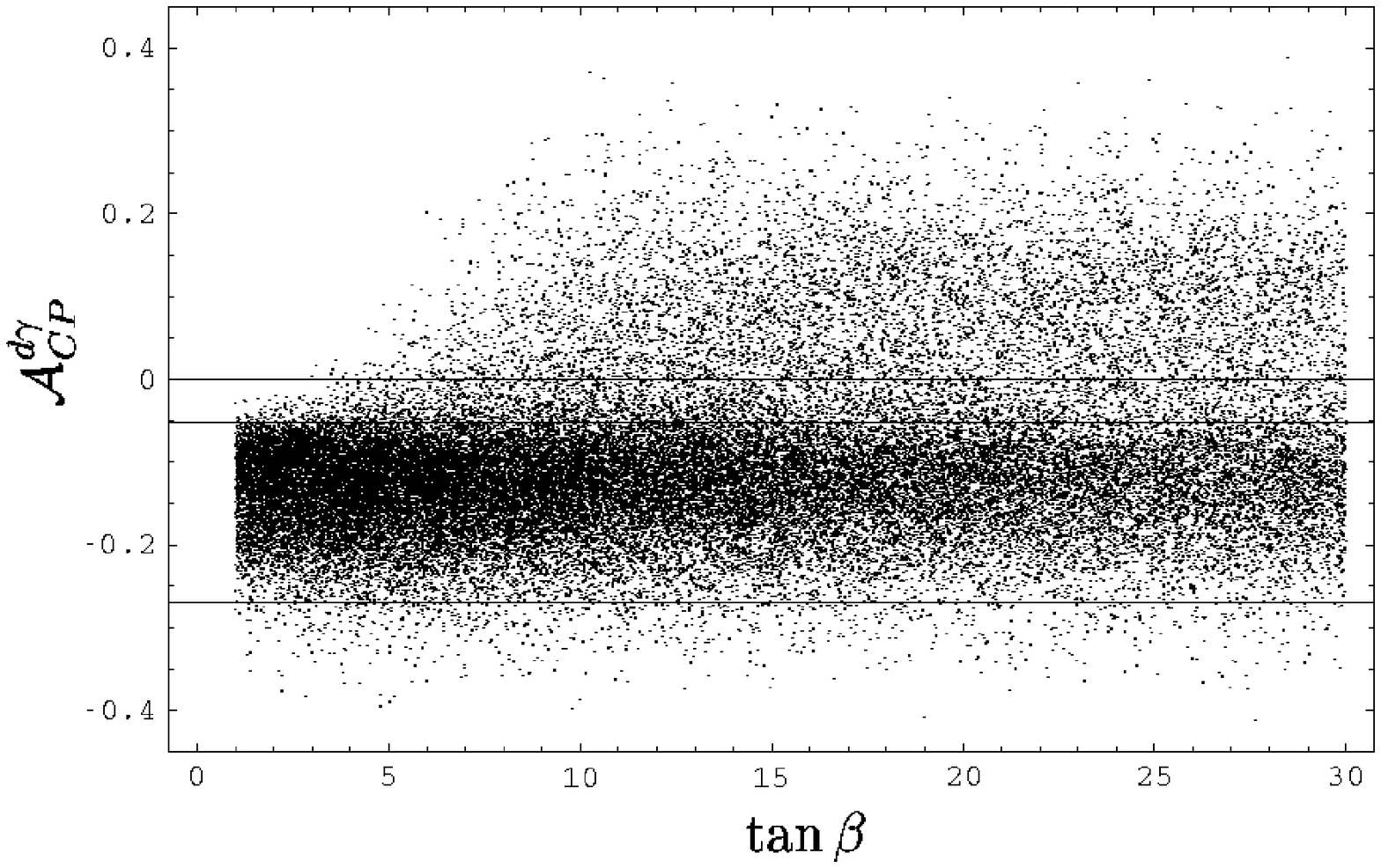}
\end{center}
\vspace*{-.5cm}
\caption{${\cal A}^{d\gamma}_{CP}$ against $\tan\beta$}
\label{tbeta}
\end{figure}

We note that our upper limit of $+40\%$ is larger than 
the maximum value of $21\%$ attained in 
\cite{plb460}. The inclusion of the SUSY phases
has joined and expanded the two phenomenological regions found in 
\cite{plb460}, allowing CP asymmetries in the continuous region
$-40\%\le {\cal A}_{CP}^{d\gamma}\le 40\%$. 
In Fig.~\ref{tbeta} we show that the large positive asymmetries can be
found anywhere in the interval $5\le \tan\beta \le 30$, 
which is the region where the EDM constraint in \cite{PRL82} is 
comfortably satisfied. 

\begin{figure}
\begin{center}
\psfrag{XXX}{${\cal A}^{s\gamma}_{CP}$}  \psfrag{YYY}
 {${\cal A}^{d\gamma}_{CP}$}
\includegraphics[width=10cm]{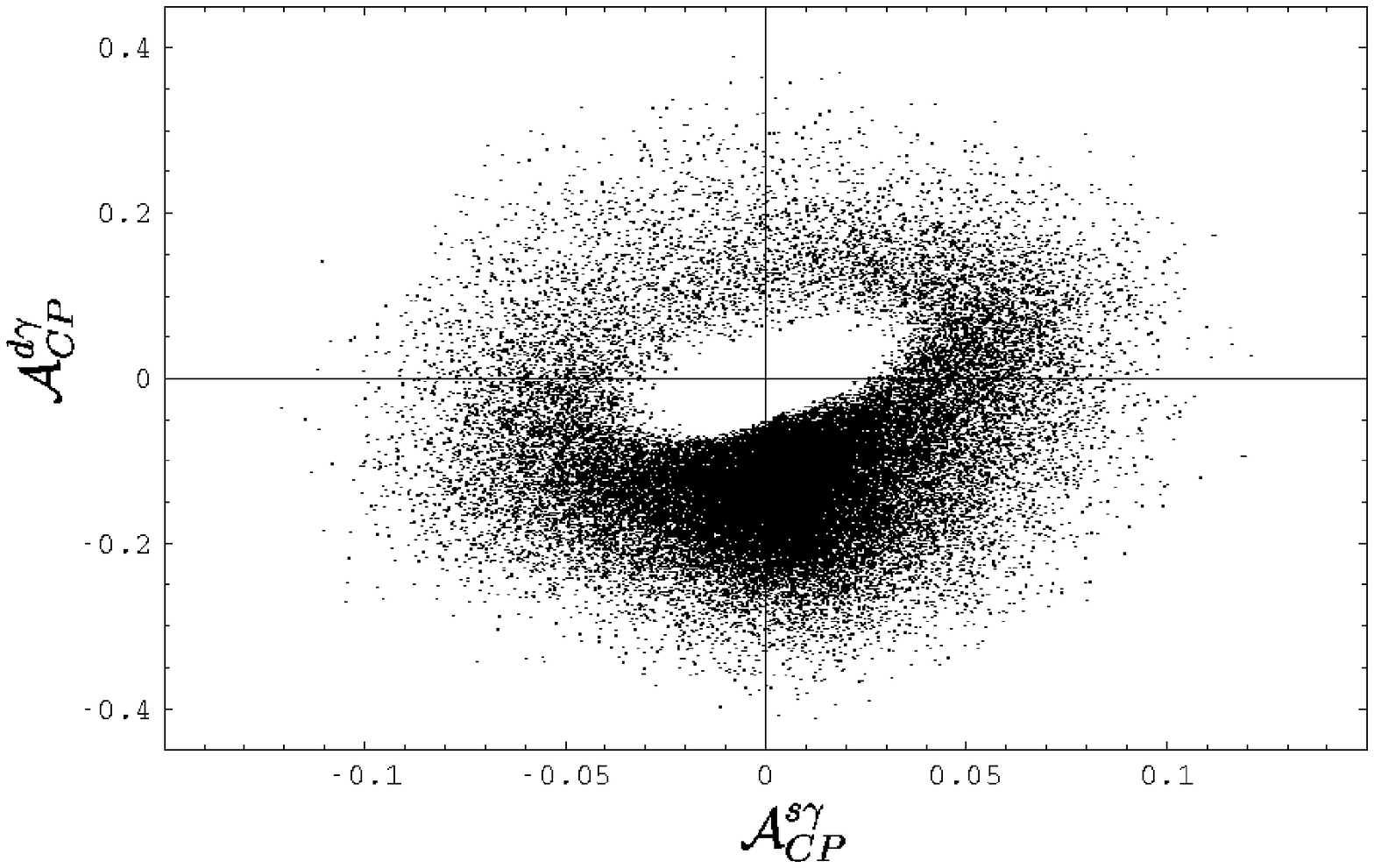}
\end{center}
\vspace*{-.5cm}
\caption{${\cal A}^{d\gamma}_{CP}$ against ${\cal A}^{s\gamma}_{CP}$}
\label{doughnut}
\end{figure}

If the signals from $b\to s\gamma$ and $b\to d\gamma$ cannot
be isolated then one must consider a combined signal.
In Fig.~\ref{doughnut} we plot ${\cal A}_{CP}^{d\gamma}$ against 
${\cal A}_{CP}^{s\gamma}$. The maximum values for
${\cal A}_{CP}^{s\gamma}$ agree with those found in \cite{baek}.
It can be seen that there is an inaccessible
region and the asymmetries can never simultaneously be
zero e.g. for ${\cal A}_{CP}^{d\gamma}\approx 0$, 
$|{\cal A}_{CP}^{s\gamma}|\ge 3\%$. This can be explained
from the fact that ${\cal A}_{CP}^{d\gamma}\approx 0$
would require $C_7$ to have a sizeable imaginary part in order to
cancel the large negative contribution from $\epsilon_d$.
The corresponding effect on ${\cal A}_{CP}^{s\gamma}$ would be to
cause a sizeable deviation from its small SM value.
Fig.~\ref{doughnut} shows that both ${\cal A}_{CP}^{s\gamma}$ and
${\cal A}_{CP}^{d\gamma}$ can have either sign, resulting
in constructive or destructive interference in eq.~(\ref{CLEOeq}).
If only the $|V_{ub}/V_{cb}|$ constraint is included in the CKM fits,
the enlarged parameter space for $\rho$ and $\eta$ allows much smaller
asymmetries for ${\cal A}_{CP}^{d,s\gamma}$. 
This is because smaller values of $\eta$ are now allowed, which reduces
the SM contribution to ${\cal A}_{CP}^{d,s\gamma}$. 
The choice of $\eta\to 0$ would correspond to points in the 
previously inaccessible region.

\begin{figure}
\begin{center}
\psfrag{XXX}{${\rm Im}(C_7)$}  \psfrag{YYY}
 {$\Delta\Gamma_d+\Delta\Gamma_s$}
\includegraphics[width=10cm]{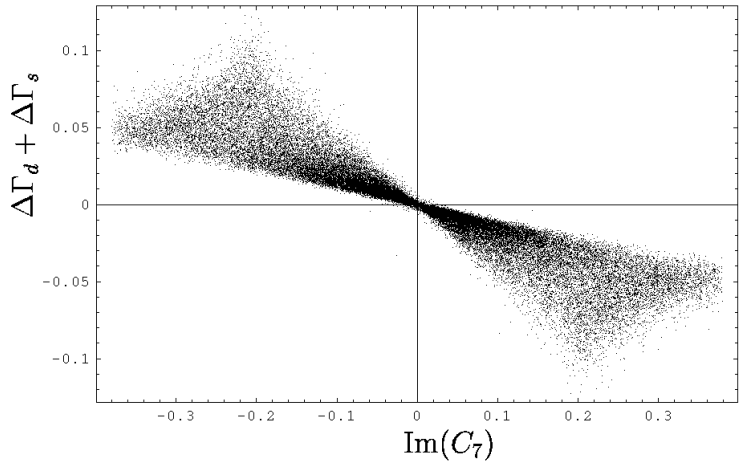}
\end{center}
\vspace*{-.5cm}
\caption{$\Delta\Gamma_d+\Delta\Gamma_s$ against ${\rm Im} (C_7)$}
\label{butterfly}
\end{figure}

In Fig.~\ref{butterfly} we plot $\Delta\Gamma_d+\Delta\Gamma_s$ 
(defined in eq.~(\ref{ACPdef})) against $\rm {Im} (C_7)$.
In the SM (as explained in Section 2) this sum would be exactly 
zero in the limit $m_s=m_d=0$ (neglecting the small
long distance contribution). From Fig.~\ref{butterfly} it can be seen 
that $\Delta\Gamma_d+\Delta\Gamma_s$ is close to 0 if $C_7$ is real, 
the slight deviation being caused by the imaginary parts of the
other Wilson coefficients. The effect of 
a non-zero $\rm{Im} (C_7)$ causes sizeable deviations from zero.

\begin{figure}
\begin{center}
\psfrag{XXX}{${\cal A}^{s\gamma}_{CP}$}  \psfrag{YYY}
 {${\cal A}^{exp}_{CP}$}
\includegraphics[width=7.5cm]{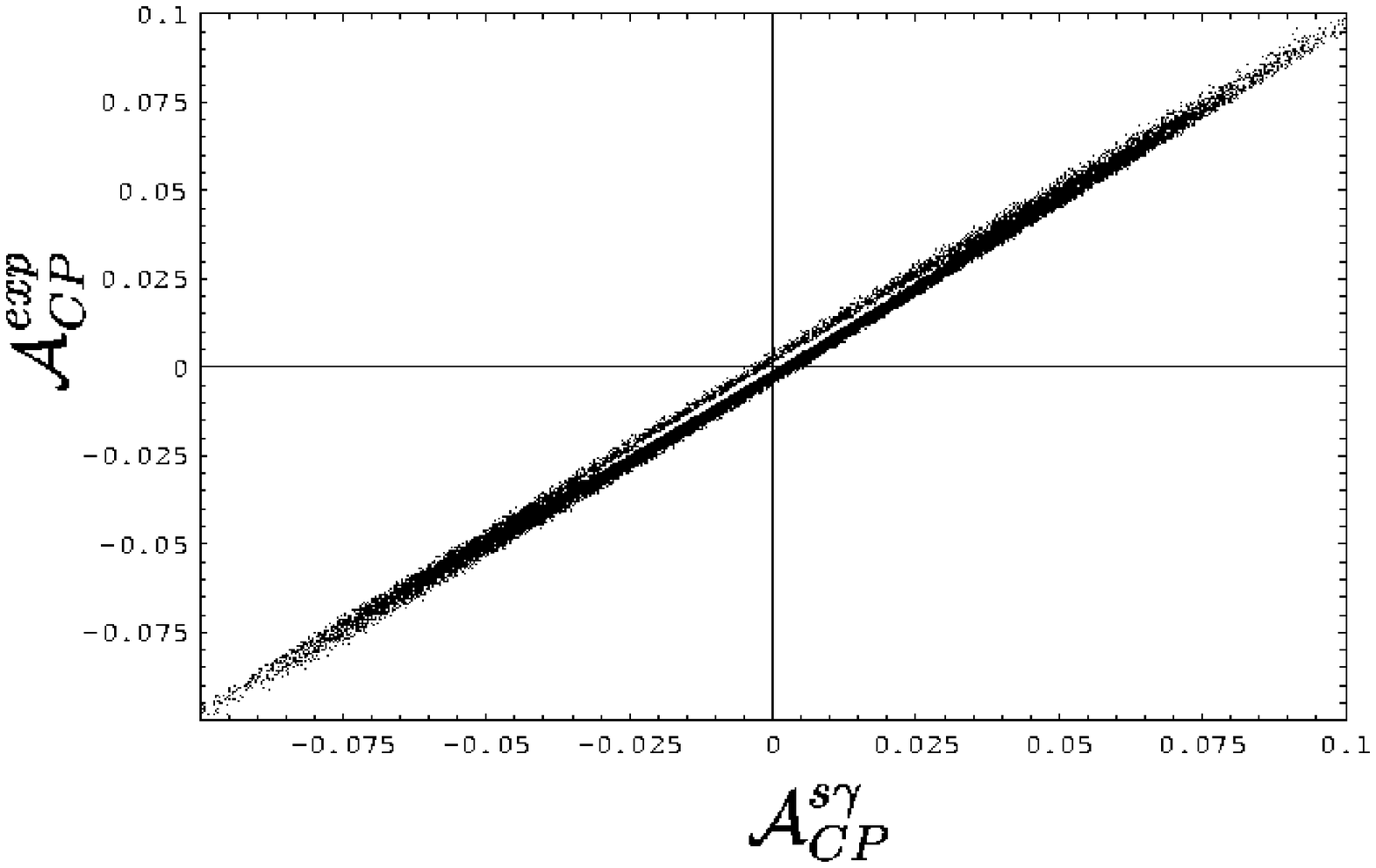}
\psfrag{XXX}{${\cal A}^{s\gamma}_{CP}$}  \psfrag{YYY}{}
\includegraphics[width=7.5cm]{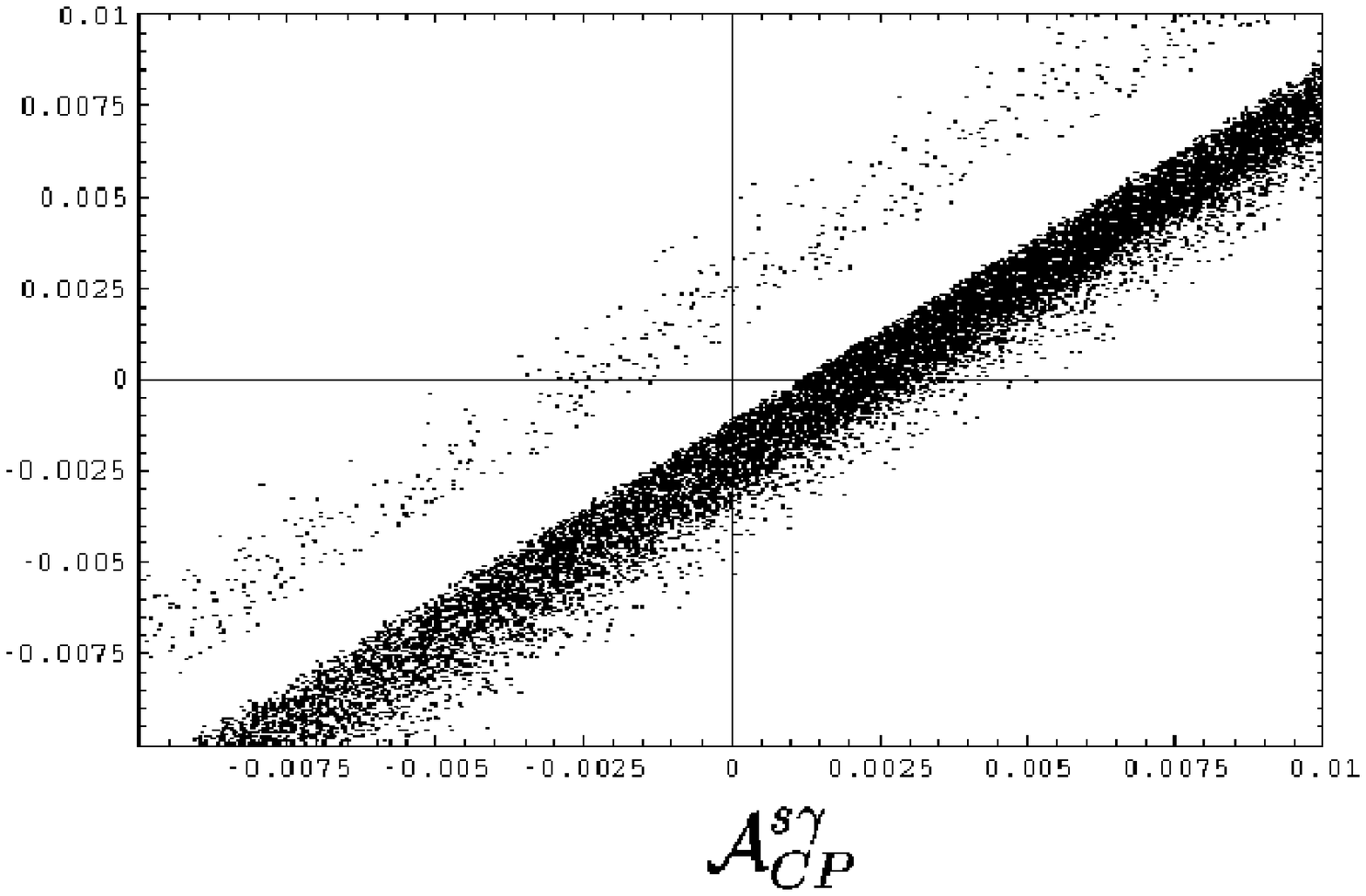}
\end{center}
\vspace*{-.5cm}
\caption{Combined asymmetry measured by CLEO $({\cal A}^{exp}_{CP})$ against 
 ${\cal A}^{s\gamma}_{CP}$}
\label{ACPCLEO}
\end{figure}

In Fig.~\ref{ACPCLEO} we plot the ${\cal A}_{CP}^{exp}$ 
(defined in eq.~(\ref{CLEOeq}))
against ${\cal A}_{CP}^{s\gamma}$. The right hand plot shows
a magnification of the area around the origin. The coefficient
of ${\cal A}_{CP}^{d\gamma}$ in eq.~(\ref{CLEOeq}) assumes that BR$(b\to d\gamma)$=
BR$(b\to s\gamma)/20$. Since this ratio of BRs is $\sim |V_{td}/V_{ts}|^2$,
which in turn is a function of the variables $\rho$ and $\eta$, we replace
the factor $1/20$ by the above ratio of CKM matrix elements.
If the contribution from ${\cal A}_{CP}^{d\gamma}$ were ignored
in eq.~(\ref{CLEOeq}), then Fig.~\ref{ACPCLEO} would be a straight 
line through the origin. The ${\cal A}_{CP}^{d\gamma}$ contribution 
broadens the line to a thin band of width $\approx 1\%$, an effect
which should be detectable at proposed higher luminosity runs of the
$B$ factories. 

Note that the width of the line is determined by the amount of
$b\to d\gamma$ admixture in the $b\to s\gamma$ sample,
eq.~(\ref{ACPdef}). In the case of the CLEO
measurement the admixture of $b\to d\gamma$ is about 2.5 times less than
the ``natural'' admixture (ratio of the branching ratios). If the
experimental analysis can be done with a natural admixture or even
a $b\to d\gamma$ enriched sample, the width of the line would 
be correspondingly
broader. Specifically, for the natural admixture the
line would be broadened by a factor of 2.5, making 
$b\to d\gamma$ a $2.5\%$ effect.
This effect is the same magnitude as the precision
attainable with an integrated luminosity of 200 fb$^{-1}$ at the
$B$ factories \cite{alex}. At this luminosity it will therefore be possible to
test the cancellation of the asymmetries as predicted by the SM.

\section{Conclusions}
We have studied the effect of supersymmetric (SUSY) CP violating phases 
on the direct CP asymmetry in the inclusive decay $B\to X_d\gamma$. 
We have performed our calculation in the effective SUSY Model, 
which allows unrestricted SUSY phases without violating the 
stringent constraints from the electron EDM. Although
such a decay contains a large background from $B\to X_s\gamma$,
it may exhibit large direct CP violation in the context of the
Standard Model (SM) and its extensions.

In the SM the CP asymmetry ${\cal A}^{d\gamma}_{CP}$
is expected to lie in the range $-5\% \le {\cal A}_{CP}\le -28\%$.
A previous analysis in the context of the MSSM in the absence of SUSY 
phases found two phenomenologically acceptable regions in SUSY parameter
space corresponding to asymmetries of $-45\% \le {\cal A}_{CP} \le -5\%$
and $2\% \le {\cal A}_{CP} \le 21\%$. The latter would constitute a
clear signal of physics beyond the SM. 
We have shown that the inclusion of phases in the SUSY breaking
parameters joins and expands these regions to allow CP asymmetries
in the continuous region $-40\% \le {\cal A}_{CP} \le 40\%$,
where the exact boundaries depend strongly on the bounds for
$C_7$ that are imposed. The largest values for the
positive asymmetry occur when the stop is lighter than 200 GeV.
Asymmetries of this magnitude are expected to be within the
reach of the $B$ factories BELLE and BaBar.

If the inclusive decay $B\to X_d\gamma$ cannot be isolated from
$B\to X_s\gamma$ then one must consider a combined signal. In the
SM the sum of the dominant short distance contributions is
identically zero (for $m_d=m_s=0$), although such a cancellation does not
occur in the effective SUSY model since $C_7$ may possess an
imaginary part. We have studied the correlation between
${\cal A}^{d\gamma}_{CP}$ and ${\cal A}^{s\gamma}_{CP}$, showing that
there may be both constructive and destructive interference.
The contribution of $B\to X_d\gamma$ to the combined signal 
${\cal A}^{exp}_{CP}$ is
suppressed by a branching ratio factor $|V_{td}/V_{ts}|^2$, 
but may be partly compensated by its potentially larger asymmetry. 
These results will be particularly important at proposed high 
luminosity runs of the $B$ factories, in which
the experimental error in the measurement of ${\cal A}^{exp}_{CP}$
is expected to reach a level comparable to the magnitude of the 
$B\to X_d\gamma$ contribution.
\vspace{20mm}
\begin{center}
{\large\bf  Acknowledgments} 
\end{center}

A.A and S.R wish to thank K.~Hagiwara for his encouragement
and the participants of the poster session at BCP4 for useful
discussions.
A.A. and S.R were supported by the Japan Society for the Promotion
of Science (JSPS).

YYK wishes to thank H.Y.Cheng 
for his encouragement and S. Baek for useful comments. 
The work of YYK was supported by the National
Science Council of R.O.C. under the Grant No. NSC-90-2811-M-002
on CP violation in $B$ meson Physics.


\newpage

\end{document}